\documentclass[jgr, draft]{agutex}
\usepackage[T1]{fontenc}
\usepackage[latin9]{inputenc}
\usepackage{units}
\usepackage{textcomp}
\usepackage{amssymb}
\usepackage{graphicx}
\usepackage{subscript}

\makeatletter
\usepackage[small]{caption}

\setkeys{Gin}{draft=false}

\authorrunninghead{ARCHER ET AL.}

\titlerunninghead{ROLE OF PRESSURE GRADIENTS}

\authoraddr{M. O. Archer,
Space \& Atmospheric Physics Group, The Blackett Laboratory, Imperial College London, Prince Consort Road, London, SW7 2AZ, UK.
(m.archer10@imperial.ac.uk)}

\authoraddr{D. L. Turner, Department of Earth, Planetary and Space Sciences, University of California, 603 Charles E. Young Drive, Los Angeles, California, CA 90095-1567, USA.
(dturner@igpp.ucla.edu)}

\authoraddr{J. P. Eastwood, Space \& Atmospheric Physics Group, The Blackett Laboratory, Imperial College London, Prince Consort Road, London, SW7 2AZ, UK.
(jonathan.eastwood@imperial.ac.uk)}

\authoraddr{T. S. Horbury, Space \& Atmospheric Physics Group, The Blackett Laboratory, Imperial College London, Prince Consort Road, London, SW7 2AZ, UK.
(t.horbury@imperial.ac.uk)}

\authoraddr{S. J. Schwartz, Space \& Atmospheric Physics Group, The Blackett Laboratory, Imperial College London, Prince Consort Road, London, SW7 2AZ, UK.
(s.schwartz@imperial.ac.uk)}

\makeatother

\begin{document}

\title{The role of pressure gradients in driving sunward magnetosheath flows
and magnetopause motion}

\authors{M. O. Archer, \altaffilmark{1}\\
D. L. Turner, \altaffilmark{2}\\
J. P. Eastwood, \altaffilmark{1}\\
T. S. Horbury,\altaffilmark{1}\\
and S. J. Schwartz\altaffilmark{1}}

\altaffiltext{1}{Blackett Laboratory, Imperial College London, London, SW7 2AZ,
UK.}

\altaffiltext{2}{Department of Earth, Planetary and Space Sciences, University
of California, Los Angeles, California, CA 90095-1567, USA.}
\begin{abstract}
While pressure balance can predict how far the magnetopause will move
in response to an upstream pressure change, it cannot determine how
fast the transient reponse will be. Using Time History of Events and
Macroscale Interactions during Substorms (THEMIS), we present multipoint
observations revealing, for the first time, strong (thermal + magnetic)
pressure gradients in the magnetosheath due to a foreshock transient,
most likely a Hot Flow Anomaly (HFA), which decreased the total pressure
upstream of the bow shock. By converting the spacecraft time series
into a spatial picture, we quantitatively show that these pressure
gradients caused the observed acceleration of the plasma, resulting
in fast sunward magnetosheath flows ahead of a localised outward distortion
of the magnetopause. The acceleratation of the magnetosheath plasma
was fast enough to keep the peak of the magnetopause bulge at approximately
the equilibrium position i.e. in pressure balance. Therefore, we show
that pressure gradients in the magnetosheath due to transient changes
in the total pressure upstream can directly drive anomalous flows
and in turn are important in transmitting information from the bow
shock to the magnetopause.
\end{abstract}
\begin{article}

\section{Introduction}

The location of the magnetopause in equilibrium is given by a balance
of pressure between the solar wind and magnetosphere, with the shocked
magnetosheath forming an interface between the two. However, while
pressure balance can determine how far the magnetopause should move
in response to changes in the pressure upstream of the bow shock,
it cannot predict how fast this motion will occur. \citet{glassmeier08}
showed that if the magnetosheath pressure changes over long enough
($\sim$5-7~min) timescales then the response of the magnetopause
can be treated quasistatically. In contrast, the solar wind dynamic
pressure can rapidly change due to, for example, tangential discontinuities
(TDs) exhibiting step changes in the density and thus dynamic pressure.
1-D magnetohydrodynamic (MHD) simulations \citep{volk74,wu93} of
the bow shock's response to such dynamic pressure drops predict that
the shock will move sunward and that a fast mode rarefaction wave
will propagate through the magnetosheath ahead of the transmitted
TD. This fast mode wave will exhibit a (thermal + magnetic) pressure
gradient across it which should weaken in time as the wave front expands.
Eventually arriving at the magnetopause, the wave communicates the
upstream pressure change to the boundary, which will subsequently
move in response. These waves have been observed due to such solar
wind dynamic pressure decreases \citep{maynard07}, however the magnetosheath
pressure gradients and how these affect the magnetosheath plasma and
in turn the magnetopause have not yet been measured due to a lack
of suitable multipoint observations in the magnetosheath.

\citet{wu93} predicted that if the decrease in dynamic pressure is
sufficently large, then the magnetosheath may experience a sunward
flow due to the rarefaction behind the rapdily expanding bow shock.
While anomalous sunward flows have previously been observed in the
magnetosheath \citep[e.g.][]{paschmann88,schwartz88,thomsen88,shue09},
their origins in general have been unclear due to a lack of simultaneous
observations upstream of the shock. It is possible that foreshock
transients may play an important role regarding anomalous magnetosheath
flows, since they too can modify the total pressure upstream. They
are kinetic phenomena which occur due to the ever changing orientation
of the interplanetary magnetic field (IMF) and solar wind conditions,
thereby changing the location and properties of the foreshock \citep[e.g.][]{eastwood05}.

A number of different types of foreshock transients have been identified
in both spacecraft observations and simulations including: hot flow
anomalies (HFAs) \citep[e.g.][]{schwartz85} caused by the interaction
of solar wind current sheets with the bow shock; spontaneous hot flow
anomalies, phenomenologically similar to HFAs but without the need
for a solar wind current sheet \citep[e.g.][]{zhang13,omidi13shfa};
foreshock cavities \citep[e.g.][]{sibeck02} caused due to an isolated
collection of field lines connected to the quasi-parallel bow shock;
and foreshock bubbles \citep{omidi10,turner13} due to the interaction
of backstreaming suprathermal ions with a discontinuity. These foreshock
transients are important because they can perturb the magnetopause
boundary, generating ultra-low frequency waves in the magnetosphere
and travelling convection vortices in the ionosphere \citep{sibeck99,eastwood08,eastwood11,jacobsen09,turner11,hartinger13}.

HFAs in particular are disruptions of the solar wind in the vicinity
of the bow shock caused by current sheets, usually TDs, interacting
with the shock \citep[e.g.][]{wang13a,wang13b}. If the solar wind
motional electric field $\mathbf{E}=-\mathbf{v}\times\mathbf{B}$
points into the TD on at least one side, ions specularly reflected
at the shock are channeled back along the current sheet \citep{burgess89,thomas91}.
The resulting hot ion population expands forming a core region of
depleted density and magnetic field and laterally drives pile up regions
and shock waves either side \citep{fuselier87,lucek04b}.

\citet{sibeck99} presented observations of a sunward plasma velocity
at the flank magnetopause boundary, which was distorted from its usual
shape into an outward bulge due to an HFA that was simultaneously
observed upstream of the shock. In contrast, \citet{jacobsen09} showed
that HFAs in the flanks can deform the magnetopause inwards and cause
fast anomalous flows. The authors explain these observations as being
due to the reduced (increased) total pressure in the HFA core (compression)
regions being transmitted through the magnetosheath and thus the magnetopause
moving in order to balance the pressure. This is broadly in agreement
with the suggestions from simulations of HFAs \citep{lin02,omidi07},
which also predict marginally sunward flows in the magnetosheath due
to the presence of the HFA. Nonetheless, the mechanism by which upstream
pressure variations due to foreshock transients are transmitted through
the bow shock and magnetosheath to the magnetopause are poorly understood.

In this paper we determine the pressure gradient in the magnetosheath
for the first time using multipoint observations, quantitatively showing
that these gradients directly drive the acceleration of the plasma
resulting in fast sunward flows and causing the subsequent motion
of the magnetopause. Through simultaneous observations upstream of
the bow shock we show that the observed gradients in the magnetosheath
were due to a foreshock transient, most likely an HFA, which reduced
the upstream total pressure.

\section{Observations}

\subsection{Magnetosheath \& Magnetopause Observations}

\begin{figure}
\begin{centering}
\includegraphics{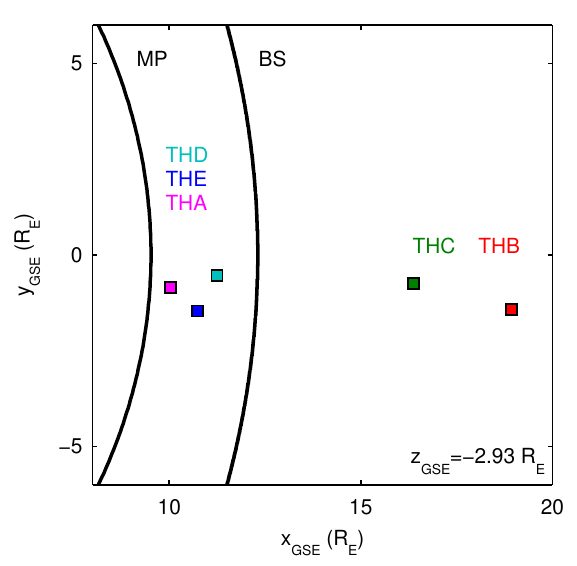}
\par\end{centering}

\caption{Positions of the five THEMIS spacecraft in GSE on 04 September 2008
at 17:32-42 UT. The \citet{shue98} model magnetopause (MP) and the
\citet{farris91} model bow shock (BS) with standoff distance set
by \citet{farrisrussell94} are also shown.\label{fig:positions}}

\end{figure}

\begin{figure}
\begin{centering}
\includegraphics[scale=0.9]{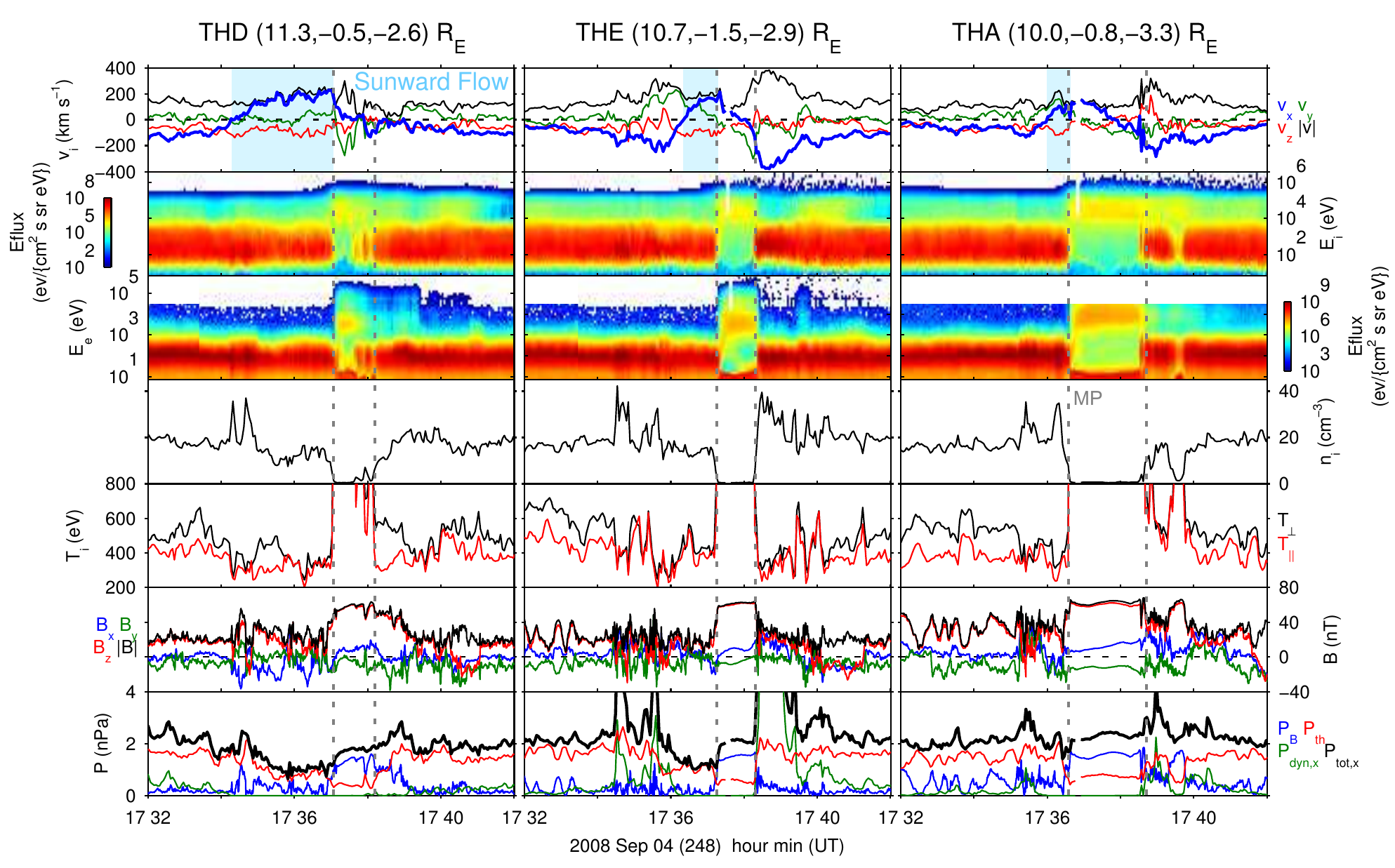}
\par\end{centering}

\caption{Magnetosheath observations from THD (left), THE (middle) and THA (right).From
top to bottom: the ion velocity components in GSE (xyz in blue, green,
red) and magnitude (black); ion energy spectrogram where the colour
scale represents the differential energy flux; electron energy spectrogram;
ion density; ion temperatures parallel (red) and perpendicular (black)
to the magnetic field; magnetic field components in GSE (xyz in blue,
green, red) and magnitude (black); and the magnetic (blue), thermal
(red), anti-sunward dynamic (green) and total anti-sunward (black)
pressures. Plasma and magnetic field data are shown at resolutions
of 3~s and 0.25~s respectively. The blue shaded regions indicate
magnetosheath flows with a sunward component and vertical grey dashed
lines highlight magnetopause crossings.\label{fig:event1-magnetosheath}}
\end{figure}

On 04 September 2008 between 17:32-42 UT, three of the THEMIS \citep{angelopoulos08}
spacecraft (THD, THE and THA) were located in the subsolar magnetosheath
separated by $\sim$1 $\mathrm{R_{E}}$ (see Figure \ref{fig:positions}).
Combined Electrostatic Analyzer (ESA) \citep{mcfadden08a} and Solid
State Telescope (SST) moments and energy spectrograms are displayed
in Figure \ref{fig:event1-magnetosheath}, revealing a transient change
in the magnetosheath ion velocity at all three spacecraft from the
regular $\sim$100 km s$^{-1}$ flow in an anti-sunward direction
to enhanced magnetosheath flows travelling sunwards (blue shaded regions).
The sunward component of the velocity was fastest ($\sim$230 km s$^{-1}$)
and longest ($\sim2\nicefrac{3}{4}$ min) at THD, furthest from Earth,
whereas it was slowest ($\sim$120 km s$^{-1}$) and shortest ($\sim$40
s) at THA, closest to Earth. Following the sunward flows, the magnetopause
passed over all three spacecraft, which subsequently had brief excursions
in the magnetosphere, before encountering the boundary again and returning
to the magnetosheath. The relative magnetopause crossing times (vertical
dashed lines in Figure \ref{fig:event1-magnetosheath}) were, however,
inconsistent with a global ``breathing'' motion of the boundary,
which would result in nested crossings due to the spacecraft's geocentric
distances.

We used minimum variance analysis (MVA) \citep[e.g.][]{sonnerup98}
of the spin-resolution Fluxgate Magnetometer (FGM) data \citep{auster08}
to determine normals for the observed magnetopause crossings, testing
the quality of the analysis via the intermediate-to-minimum eigenvalue
ratio test ($\lambda_{int}/\lambda_{min}\gtrsim10$ implying a reliable
normal) as well as the sensitivity of the resulting normal to different
time intervals centred on the boundary. For the inbound crossings,
the THD observations provided the most reliable normal $\mathbf{n}=$(0.70,-0.67,-0.23)
in GSE coordinates which is deflected, predominantly towards the -y
GSE direction, by $36^{\circ}$ from the expected orientation of the
normal $\mathbf{N}$ from the \citet{shue98} model magnetopause.
For all spacecraft pairs we used the two spacecraft timing method
\citep[e.g.][]{schwartz98} to estimate the magnetopause speed along
the normal, yielding an average value $v_{n}=122\pm7$ km s\textsuperscript{-1}:
much faster than typical motions though within the range of those
previously observed \citep[e.g.][]{plaschke09}. The orientation of
the magnetopause crossing indicates that there was a localised outwards
distortion of the magnetopause moving across the model boundary. The
transit velocity of this bulge $\mathbf{v}_{trans}$, given by \citep{schwartz00}

\begin{equation}
\mathbf{v}_{trans}=\frac{v_{n}}{\sin^{2}\theta_{Nn}}\left(\mathbf{n}-\cos\theta_{Nn}\mathbf{N}\right)
\end{equation}
where $\theta_{Nn}$ is the angle between the two normals, was found
to be (-19,-205,12) km s$^{-1}$ in GSE coordinates. While no reliable
magnetopause normals could be obtained from MVA for the outbound crossings,
we estimated the trailing edge's orientation by three spacecraft timing
\citep{horbury01b} using the transit velocity of the leading edge
$\mathbf{v}_{trans}$. This resulted in an outbound normal deflected,
predominantly towards the +y GSE direction, by 29\textdegree{} from
the model boundary. Therefore, the magnetopause was locally distorted
into an outward bulge.

The bottom panels of Figure \ref{fig:event1-magnetosheath} show the
combined isotropic ion and electron thermal pressure $P_{th}$, the
magnetic pressure $P_{B}$ and the anti-sunward dynamic pressure $P_{dyn,x}$
(for only those intervals in which the flow was antisunward) as well
as the sum of these, the total pressure $P_{tot,x}$. All spacecraft
observed a decrease in the total pressure (from the background value
of $\sim$2~nPa) during the event, first observed at THD and followed
$\sim$55~s later by THA and THE in quick succession. This decrease
was greatest ($\sim$1.2 nPa) at THD and only marginal ($\sim$0.3
nPa) at THA. Therefore there was a pressure gradient through the magnetosheath,
driving the sunward flows and outward magnetopause motion as we show
in section \ref{sec:Analysis}. Next we investigate the origin of
this pressure gradient through observations upstream of the bow shock.

\subsection{Solar Wind \& Foreshock Observations}

\begin{figure}
\begin{centering}
\includegraphics{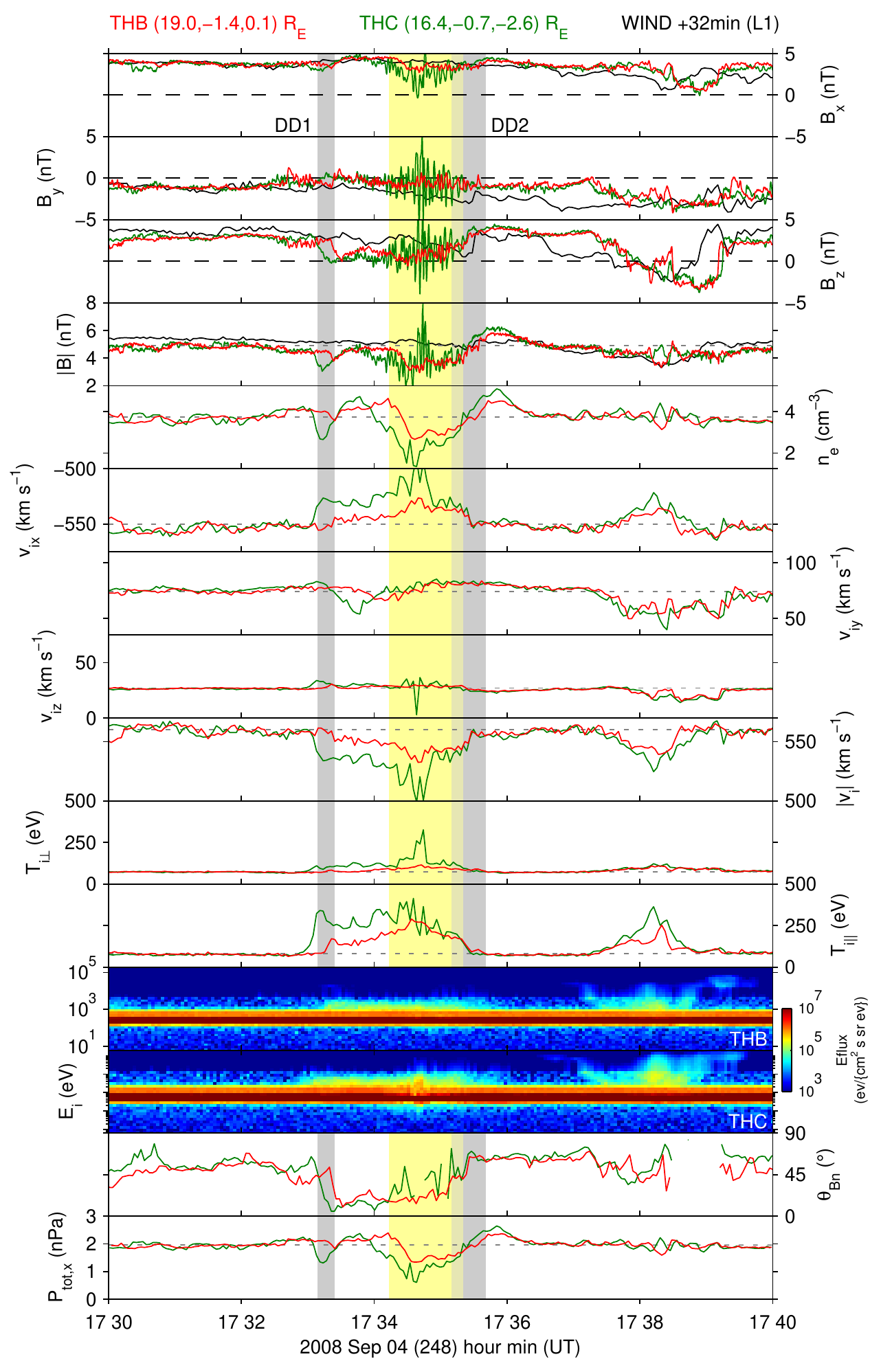}
\par\end{centering}

\caption{Solar wind/foreshock observations from THB (red), THC (green) and
WIND (black), with the latter lagged by 32 min. From top to bottom:
x,y \& z components of the magnetic field in GSE; magnetic field strength;
electron density; x,y \& z components of the ion velocity in GSE;
magnitude of the ion velocity; ion temperatures perpendicular and
parallel to the magnetic field; ion energy spectrograms; electron
energy spectrograms; magnetic field - shock normal angle $\theta_{Bn}$
magnetically connected to the spacecraft; and the total anti-sunward
pressure. Plasma and magnetic field data are shown at resolutions
of 3~s and 0.25~s respectively, while $\theta_{Bn}$ is calculated
using 3~s resolution magnetic field data. Shaded areas highlight
two directional discontinuities (DD1 \& DD2) in grey and a region
of depleted density and magnetic field in yellow.\label{fig:event1-upstream}}
\end{figure}

Figure \ref{fig:event1-upstream} shows simultaneous observations
upstream of the bow shock from THB (red) and THC (green) along with
Magnetic Field Investigation data \citep{lepping95} from the WIND
spacecraft (black) near L1, which has been lagged by 32 min (plasma
data is not shown due to its low time resolution of 97 s which revealed
no strong variations). The time lag was obtained by matching up the
magnetic field signatures between WIND and the THEMIS spacecraft.
We highlight (grey areas) two directional discontinuities, denoted
DD1 and DD2, between which the IMF was radial and backstreaming suprathermal
ions (see ion spectrograms) typical of the ion foreshock \citep[e.g.][]{eastwood05}
were observed. These suprathermal populations caused the observed
increases in parallel ion temperature moments (over the entire distributions)
during this period.

On the downstream edge of DD2, a region of depleted density and magnetic
field was observed by both THEMIS spacecraft (yellow area) with compressions
either side. These correlated variations, which were not present in
the almost steady pristine solar wind, are indicative of foreshock
transients \citep{fairfield90}. Within the depleted core, ``3 s
waves'' \citep{le92} (almost circularly polarised waves that are
right-handed in the spacecraft frame) were observed which typically
occur in the foreshock under high solar wind plasma $\beta$ conditions
($\beta\sim$5 here); the solar wind was slowed; and the ion temperatures
moments (over the entire distributions) marginally increased both
parallel and perpendicular to the magnetic field. In addition the
ion distributions (not shown) were more diffuse compared to the intermediate
distributions observed outside of this region, though the solar wind
beam persisted throughout. Note that these variations were observed
simultaneously at both THEMIS spacecraft but were all larger at THC,
closer to the bow shock.

Since MVA was poorly conditioned ($\lambda_{int}/\lambda_{min}\sim1$),
we determined the orientation of DD2 using a constrained two spacecraft
timing method between THB and WIND. The normal $\mathbf{n}_{DD2}$
was constructed using the regular two spacecraft timing method ($\left\{ \mathbf{r}_{\alpha\beta}-\mathbf{v}_{sw}t_{\alpha\beta}\right\} \cdot\mathbf{n}_{DD2}=0$
where $\mathbf{r}_{\alpha\beta}$ is the spacecraft separation vector
and $t_{\alpha\beta}$ the relative timing between the spacecraft)
and by requiring it be perpendicular to the maximum variance direction
($\mathbf{e}_{max}\cdot\mathbf{n}_{DD2}=0$), which was better defined
than the minimum ($\lambda_{max}/\lambda_{int}\sim24$). The computed
normal was found to be almost entirely in the GSE y direction: (0.09,0.99,0.10)
in GSE coordinates. This was close (within $\sim12^{\circ}$) to the
theoretical normal for a TD given by the cross product of the upstream
and downstream magnetic fields \citep[e.g.][]{knetter04}.

Considering the thermal, magnetic and dynamic pressures upstream of
the shock (bottom panel of Figure \ref{fig:event1-upstream}), it
was found that the total pressure associated with the core of this
transient was dominated by the dynamic pressure and this varied chiefly
due to the density changes rather than the velocity. The total pressure
at THC, closest to the bow shock, decreased (from its ambient value
of $\sim$2.0 nPa) by $\sim$1.4 nPa in the depeleted core, with increases
of $\sim$0.3 nPa and $\sim$0.7 nPa before and after respectively;
thus the total pressure upstream of the bow shock was modified due
to the presence of the transient.

This foreshock transient cannot be explained as a Foreshock Bubble
since it satisfies none of the \citet{turner13} criteria for Foreshock
Bubble identification. This transient could be a Foreshock Cavity,
since there was a localised region of plasma connected to the quasi-parallel
shock \citep{schwartz06} between DD1 and DD2. However, the transient
formed at DD2 (most clear at THB) rather than filling the space between
the two discontinuities, unlike a typical Foreshock Cavity. In contrast,
the event did satisfy all the \citet{schwartz00} conditions for HFA
identification: the electric field pointed into the current sheet
on the upstream side and the discontinuity normal was almost perpendicular
to the Sun-Earth line. In addition there was very little change in
the magnetic field strength across the discontinuity and quasi-perpendicular
bow shock conditions were present on the upstream side, which are
typical for HFAs \citep{schwartz00}. Further evidence towards an
HFA include the depleted core being displaced towards the side of
the discontinuity magnetically connected to the quasi-parallel bow
shock \citep{omidi07,zhang10,wang13b}; the existence of compressions
either side of this core region \citep[e.g.][]{schwartz88}; and the
differences between the two THEMIS spacecraft, showing simultaneously
weaker variations further upstream, in agreement with the locality
of HFAs to the bow shock. However, the plasma moments showed no large
deflections in the ion velocity nor significant heating of the thermal
ion plasma, the typical signatures of an HFA. The observations were
very similar to the ``proto-HFA'' presented by \citet{zhang10},
in which simultaneous observations closer to the bow shock revealed
stronger HFA signatures including significant heating and shocks i.e.
the signatures of the ``proto-HFA'' were observations taken further
upstream of an HFA at the bow shock. On the other hand, \citet{omidi13}
have shown through 2-D kinetic hybrid simulations that a succession
of two discontinuities resulting in a localised foreshock between
them, similar to the case here, can lead to somewhat similar correlated
density and magnetic field variations. While we conclude this foreshock
transient was most likely an HFA (and will refer to it as such henceforth),
we cannot unambiguously determine so. Nonetheless, it is the fact
that the pressure upstream of the bow shock was altered by the transient
which is important here, not its specific nature.

\section{Analysis\label{sec:Analysis}}

\begin{figure}
\begin{centering}
\includegraphics{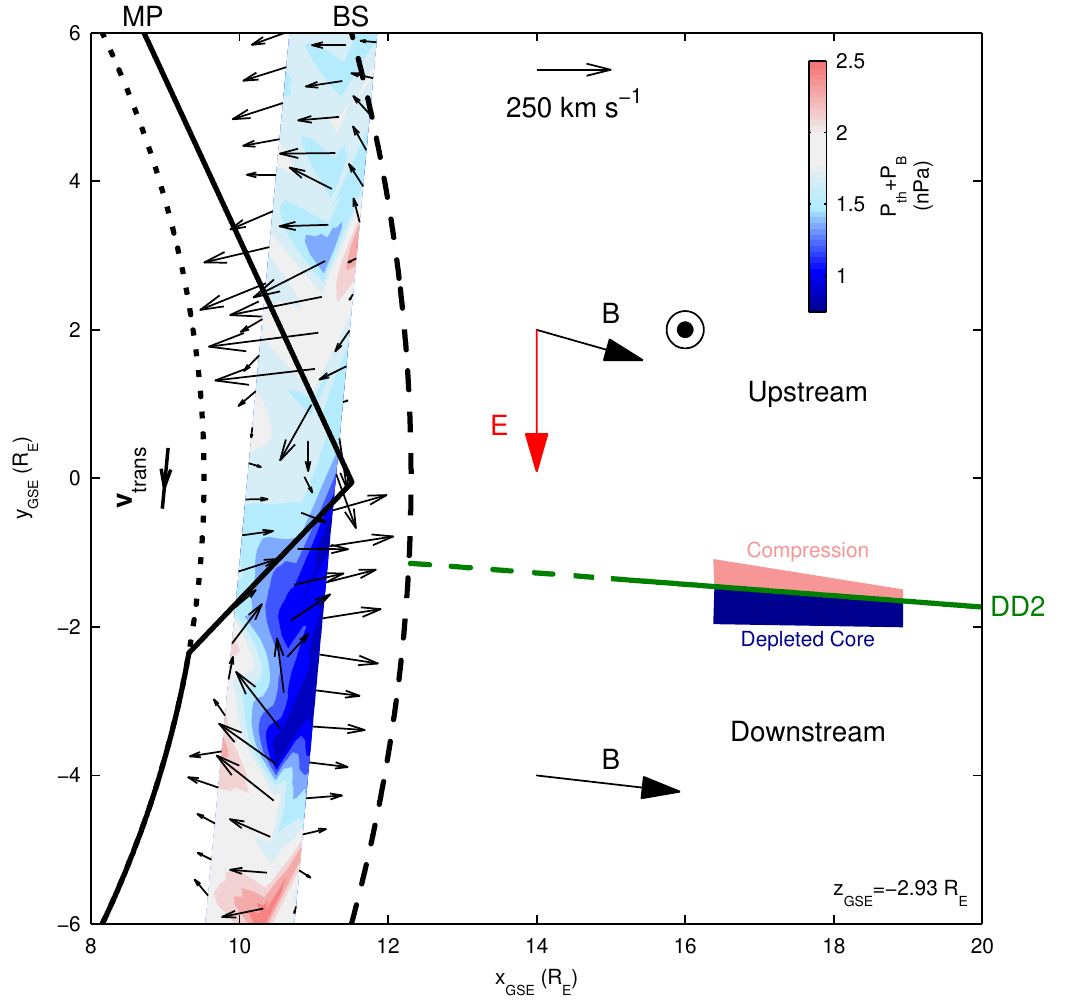}
\par\end{centering}

\caption{Snapshot of the event at 17:37:00 UT in the x-y GSE frame. The observed
magnetopause (MP) deformation (solid black) combined with the \citet{shue98}
model (dotted black), the \citet{farris91} model bow shock (BS, dashed
black) and the current sheet DD2 (green) are shown. The IMF and motional
electric field either side of DD2 are given by the black and red arrows
respectively (note that on the downstream side the electric field
is negligible due to the radial IMF). We display the observed flow
pattern (black arrows), the transit speed $\mathbf{v}_{trans}$ of
the magnetopause bulge, and contours of the thermal + magnetic pressure
(colour scale). The observed depleted core and compression regions
of the transient are indicated either side of DD2, with colours representative
of the total pressure in each. \label{fig:event1-schematic}}
\end{figure}

To determine whether the HFA was the cause of the sunward flows and
magnetopause motions, we piece together the observations from all
five THEMIS spacecraft. We assume that both the observed HFA and magnetopause
disturbance time series were due to the structures simply passing
over the spacecraft and not intrinsically temporal variations. This
assumption leads to a consistent framework for this event, except
for immediately preceding the pressure reduction where the observed
density enhancements varied significantly between spacecraft meaning
that there was substructure in either space or time. It is possible
to turn the observed time series into the spatial picture shown in
Figure \ref{fig:event1-schematic}, a snapshot of the event in the
GSE frame at 17:37 UT. Since the spacecraft separation vectors, observed
motions and estimated normals all have relatively small components
in the GSE z direction, we can limit our spatial picture to the GSE
x-y plane for simplicity. For the shape of the magnetopause (solid
black line), we combine the \citet{shue98} model boundary (dotted
black line) with the determined outward bulge. Using the transit velocity
of the bulge $\mathbf{v}_{trans}$, we arrive at the spatial variations
in the (thermal + magnetic) pressure and velocity fields in the magnetosheath
along the spacecraft tracks. In the case of the pressure, this has
been interpolated between the three spacecraft tracks yielding the
pressure contours shown in Figure \ref{fig:event1-schematic}.

DD2 (green) is indicated upstream of the \citet{farris91} model bow
shock (black dashed line), showing good agreement between the intersection
of the (assumed planar) current sheet with the model shock and the
peak location of the outward bulge of the magnetopause. By using the
total pressure observed in the HFA core from THC, we estimate through
pressure balance that the magnetopause should have locally moved out
to a GSE x location of 11.3 R\textsubscript{E} in the plane displayed
in Figure \ref{fig:event1-schematic}, which agrees well with the
intersection of the (assumed planar) leading and trailing edges of
the boundary disturbance at 11.5 R\textsubscript{E}.

Figure \ref{fig:event1-schematic} shows a region of reduced (thermal
+ magnetic) pressure in the magnetosheath, spanning $\sim$4 $\mathrm{R_{E}}$
transverse to the bow shock, ahead of the leading edge of the magnetopause
bulge. This was due to the presence of the HFA upstream of the bow
shock, in particular its depleted core on the downstream edge of DD2
with decreased total pressure. The existence of a localised region
of reduced pressure means that there were substantial pressure gradients
in the magnetosheath. We wish to determine quantitatively whether
these gradients can account for the observed accelerations of the
magnetosheath flow to a sunward direction. We calculate the local
(Eulerian) acceleration $\partial\mathbf{v}/\partial t$ observed
by the THEMIS spacecraft from a linear fit to the velocity time series
(periods indicated by the horizontal blue bars at the top of Figure
\ref{fig:event1-magnetosheath}) and then compare this to that expected
from MHD theory. The MHD momentum equation (neglecting the magnetic
tension force since $\beta\sim$3-8 within the sunward flows) is given
by

\begin{equation}
\frac{\partial\mathbf{v}}{\partial t}=-\frac{1}{\rho}\nabla\left(P_{th}+P_{B}\right)-\mathbf{v}\cdot\nabla\mathbf{v}\label{eq:MHD-momentum}
\end{equation}
Therefore, using the pressure and velocity fields derived from the
multiple spacecraft, it is possible to calculate the expected local
acceleration $\partial\mathbf{v}/\partial t$ at the spacecraft locations
in the GSE z=-2.93~R\textsubscript{E} plane, as shown in Figure
\ref{fig:event1-schematic}. THD, furthest from Earth, observed a
sunward acceleration of (2.9,-0.1) km s$^{-2}$ in this plane. Using
the right hand side of Equation \ref{eq:MHD-momentum}, we arrive
at an expected acceleration in this direction (primarily due to the
pressure gradient since the advective term was small) of 2.7 km s$^{-2}$.
Similarly THA, closest to Earth, observed an acceleration of (4.6,3.1)
km s\textsuperscript{-2}, which has a magnitude of 5.5 km s\textsuperscript{-2},
and we compute the expected acceleration due to the pressure gradient
in this direction to be 5.6 km s\textsuperscript{-2}. The sunward
acceleration at THE varied from a value similar to that at THA to
one similar to THD. The observed sunward accelerations of the magnetosheath
plasma were therefore in excellent agreement with that predicted by
MHD due to the determined pressure gradients. These are different
at the three spacecraft due to the localised nature of the pressure
decrease in the magnetosheath, meaning that the orientation of the
pressure fronts (and thus the pressure gradients) are highly dependent
on location, as can be seen in Figure \ref{fig:event1-schematic}.

Fast anti-sunward flows were also observed (most notably by both THE
and THA) on the trailing edge of the magnetopause distortion, but
again before the pressure decrease at THE. Here they are also likely
due to the pressure gradient force: the total pressure upstream of
the bow shock was enhanced in the HFA's compression regions and was
reduced within the outward bulge of the magnetopause. Slight thermal
pressure increases were observed at THE during the anti-sunwards flows
($\sim$0.8~nPa and $\sim$0.5~nPa respectively), however similar
increases at THA were obscured due to oscillations of the boundary.
The observed jets, which enhance the total pressure on the magnetopause
chiefly due to their dynamic pressure (Figure \ref{fig:event1-magnetosheath}),
could cause inward distortions of the magnetopause \citep[e.g.][]{shue09}
as have been observed due to HFAs previously \citep{jacobsen09}.

\section{Conclusions}

We have presented a case study of fast sunward magnetosheath flows
followed by motions of the magnetopause. Simultaneous observations
upstream of the bow shock revealed that these were due to a foreshock
transient, most likely a Hot Flow Anomaly (HFA), which had an associated
decrease in total pressure. By converting the spacecraft time series
into a spatial picture, we have directly shown the pressure gradients
in the magnetosheath due to this pressure decrease for the first time.
These pressure gradients drive the acceleration that resulted in the
anamalous flow patterns, transmitting the information of the upstream
pressure change through the magnetosheath to the magnetopause. In
turn the boundary is no longer in pressure balance and moves, causing
a localised outwards distortion of the magnetopause. Furthermore,
the acceleratation of the magnetosheath plasma was fast enough to
keep the peak of the magnetopause bulge at approximately the equilibrium
position i.e. in pressure balance.

Previous studies have shown that the reduced pressure of HFA cores
can be transmitted through to the magnetosheath \citep{eastwood08}
and subsequently cause magnetopause motions \citep{sibeck99,jacobsen09}
via pressure balance. Now, thanks to the geometry of the HFA and the
configuration of the THEMIS spacecraft, we have shown the role of
the pressure gradient force in directly driving the magnetosheath
flow, with the calculated pressure gradients agreeing quantitatively
with the measured acceleration of plasma. Previous 2-D hybrid simulations
of HFAs \citep{lin02,omidi07} resulted in only very small sunward
components of the magnetosheath velocity. Observations \citep{eastwood08}
have shown that not all HFAs cause sunward flows in the magnetosheath,
though they do cause some flow deflections. The reason for such fast
sunward flows observed here is likely due to the amplitude of the
upstream pressure decrease, as per the 1-D MHD theory of \citet{wu93}.
Therefore, it is important to understand the factors which control
the pressure variations which develop/evolve in foreshock transients.

We have shown the role that magnetosheath pressure gradients play
in transmitting information about upstream pressure variations, in
this case due to a foreshock transient. However such variations can
originate from other types of transients \citep[e.g.][]{sibeck02,omidi10}
as well as from the solar wind \citep[e.g.][]{potemra89}. Whether
the flow is accelerated to become sunward will depend not only on
the strength of the gradients but also on the location in the magnetosheath,
with sunward flows theoretically being more easily achieved close
to the magnetopause due to the reduced magnetosheath velocities. Further
multipoint observations in the magnetosheath could therefore allow
us to better understand the pressure gradients which form due to a
number of different phenomena under different conditions and the effect
these have on driving anomalous magnetosheath flows and magnetopause
motions.
\begin{acknowledgments}
M. O. Archer would like to thank H. Hietala for helpful discussions.
This research at Imperial College London was funded by STFC grants
ST/I505713/1, ST/K001051/1 and ST/G00725X/1. D. L. Turner is thankful
for funding from NASA (THEMIS mission and grant NNX14AC16G). We acknowledge
NASA contract NAS5-02099 and V. Angelopoulos for use of data from
the THEMIS Mission, specifically C. W. Carlson and J. P. McFadden
for use of ESA data; D. Larson and R. P. Lin for use of SST data;
and K. H. Glassmeier, U. Auster and W. Baumjohann for the use of FGM
data provided under the lead of the Technical University of Braunschweig
and with financial support through the German Ministry for Economy
and Technology and the German Center for Aviation and Space (DLR)
under contract 50 OC 0302. Finally, we acknowledge A. Szabo and K.
Ogilvie for WIND magnetic field and plasma data.
\end{acknowledgments}

\bibliographystyle{agu}
\bibliography{sunwardflows}

\begin{thebibliography}{48}
\providecommand{\natexlab}[1]{#1}
\expandafter\ifx\csname urlstyle\endcsname\relax
  \providecommand{\doi}[1]{doi:\discretionary{}{}{}#1}\else
  \providecommand{\doi}{doi:\discretionary{}{}{}\begingroup
  \urlstyle{rm}\Url}\fi

\bibitem[{\textit{Angelopoulos}(2008)}]{angelopoulos08}
Angelopoulos, V., The {THEMIS} mission, \textit{Space Sci. Rev.}, \textit{141},
  5--34, \doi{10.1007/s11214-008-9336-1}, 2008.

\bibitem[{\textit{Auster et~al.}(2008)}]{auster08}
Auster, H.~U., et~al., The {THEMIS} fluxgate magnetometer, \textit{Space Sci.
  Rev.}, \textit{141}, 235--264, \doi{10.1007/s11214-008-9365-9}, 2008.

\bibitem[{\textit{Burgess}(1989)}]{burgess89}
Burgess, D., On the effect of a tangential discontinuity on ions specularly
  reflected at an oblique shock, \textit{J. Geophys. Res.}, \textit{94},
  472--478, \doi{10.1029/JA094iA01p00472}, 1989.

\bibitem[{\textit{Eastwood et~al.}(2005)\textit{Eastwood, Lucek, Mazelle,
  Meziane, Narita, Pickett, and Treumann}}]{eastwood05}
Eastwood, J.~P., E.~A. Lucek, C.~Mazelle, K.~Meziane, Y.~Narita, J.~Pickett,
  and R.~A. Treumann, The foreshock, \textit{Space Science Reviews},
  \textit{118}, 41--94, \doi{10.1007/s11214-005-3824-3}, 2005.

\bibitem[{\textit{Eastwood et~al.}(2011)\textit{Eastwood, Schwartz, Horbury,
  Carr, Glassmeier, Richter, Koenders, Plaschke, and Wild}}]{eastwood11}
Eastwood, J.~P., S.~J. Schwartz, T.~S. Horbury, C.~M. Carr, K.-H. Glassmeier,
  I.~Richter, C.~Koenders, F.~Plaschke, and J.~A. Wild, Transient {P}c3 wave
  activity generated by a hot flow anomaly: {C}luster, {R}osetta, and
  ground-based observations, \textit{J. Geophys. Res.}, \textit{116}, A08,224,
  \doi{10.1029/2011JA016467}, 2011.

\bibitem[{\textit{Eastwood et~al.}(2008)}]{eastwood08}
Eastwood, J.~P., et~al., T{HEMIS} observations of a hot flow anomaly: solar
  wind, magnetosheath, and ground-based measurements, \textit{Geophys. Res.
  Lett.}, \textit{35}, L17S03, \doi{10.1029/2008GL033475}, 2008.

\bibitem[{\textit{Fairfield et~al.}(1990)\textit{Fairfield, Baumjohann,
  Paschmann, L\"{u}hr, and Sibeck}}]{fairfield90}
Fairfield, D.~H., W.~Baumjohann, G.~Paschmann, H.~L\"{u}hr, and D.~G. Sibeck,
  Upstream pressure variations associated with the bow shock and their effects
  on the magnetosphere, \textit{J. Geophys. Res.}, \textit{95}, 3773--3786,
  \doi{10.1029/JA095iA04p03773}, 1990.

\bibitem[{\textit{Farris and Russell}(1994)}]{farrisrussell94}
Farris, M.~H., and C.~T. Russell, Determining the standoff distance of the bow
  shock: {M}ach number dependence and use of models, \textit{J. Geophys. Res.},
  \textit{99}, 17,681--17,689, \doi{10.1029/94JA01020}, 1994.

\bibitem[{\textit{Farris et~al.}(1991)\textit{Farris, Petrinec, and
  Russell}}]{farris91}
Farris, M.~H., S.~M. Petrinec, and C.~T. Russell, The thickness of the
  magnetosheath: Constraints on the polytropic index, \textit{Geophys. Res.
  Lett.}, \textit{18}, 1821--1824, \doi{10.1029/91GL02090}, 1991.

\bibitem[{\textit{Fuselier et~al.}(1987)\textit{Fuselier, Thomsen, Gosling,
  Bame, Russell, and Mellott}}]{fuselier87}
Fuselier, S.~A., M.~F. Thomsen, J.~T. Gosling, S.~J. Bame, C.~T. Russell, and
  M.~M. Mellott, Fast shocks at the edges of hot diamagnetic cavities upstream
  from the {E}arth's bow shock, \textit{J. Geophys. Res.}, \textit{92},
  3187--3194, \doi{10.1029/JA092iA04p03187}, 1987.

\bibitem[{\textit{Glassmeier et~al.}(2008)}]{glassmeier08}
Glassmeier, K.-H., et~al., Magnetospheric quasi-static response to the dynamic
  magnetosheath: {A} {THEMIS} case study, \textit{Geophys. Res. Lett.},
  \textit{35}, L17S01, \doi{10.1029/2008GL033469}, 2008.

\bibitem[{\textit{Hartinger et~al.}(2013)\textit{Hartinger, Turner, Plaschke,
  Angelopoulos, and Singer}}]{hartinger13}
Hartinger, M.~D., D.~L. Turner, F.~Plaschke, V.~Angelopoulos, and H.~Singer,
  The role of transient ion foreshock phenomena in driving {P}c5 {ULF} wave
  activity, \textit{J. Geophys. Res.}, \textit{118}, 299--312,
  \doi{10.1029/2012JA018349}, 2013.

\bibitem[{\textit{Horbury et~al.}(2001)\textit{Horbury, Burgess, Fr\"{a}nz, and
  Owen}}]{horbury01b}
Horbury, T.~S., D.~Burgess, M.~Fr\"{a}nz, and C.~J. Owen, Three spacecraft
  observations of solar wind discontinuities, \textit{Geophys. Res. Lett.},
  \textit{28}, 677--680, \doi{10.1029/2000GL000121}, 2001.

\bibitem[{\textit{Jacobsen et~al.}(2009)}]{jacobsen09}
Jacobsen, K.~S., et~al., {THEMIS} observations of extreme magnetopause motion
  caused by a hot flow anomaly, \textit{J. Geophys. Res.}, \textit{114},
  A08,210, \doi{10.1029/2008JA013873}, 2009.

\bibitem[{\textit{Knetter et~al.}(2004)\textit{Knetter, Neubauer, Horbury, and
  Balogh}}]{knetter04}
Knetter, T., F.~M. Neubauer, T.~Horbury, and A.~Balogh, Four-point
  discontinuity observations using {C}luster magnetic field data: A statistical
  survey, \textit{J. Geophys. Res.}, \textit{109}, A06,102,
  \doi{doi:10.1029/2003JA010099}, 2004.

\bibitem[{\textit{Le et~al.}(1992)\textit{Le, Russell, Thomsen, and
  Gosling}}]{le92}
Le, G., C.~T. Russell, M.~F. Thomsen, and J.~T. Gosling, Observations of a new
  class of upstream waves with periods near 3 seconds, \textit{J. Geophys.
  Res.}, \textit{97}, 2917--2925, \doi{10.1029/91JA02707}, 1992.

\bibitem[{\textit{Lepping et~al.}(1995)}]{lepping95}
Lepping, R.~P., et~al., The {WIND} magnetic field investigation, \textit{Space
  Sci. Rev.}, \textit{71}, 207--229, \doi{10.1007/BF00751330}, 1995.

\bibitem[{\textit{Lin}(2002)}]{lin02}
Lin, Y., Global hybrid simulation of hot flow anomalies near the bow shock and
  in the magnetosheath, \textit{Planet. Space Sci.}, \textit{50}, 577--591,
  \doi{10.1016/S0032-0633(02)00037-5}, 2002.

\bibitem[{\textit{Lucek et~al.}(2004)\textit{Lucek, Horbury, Balogh, Dandouras,
  and R\`{e}me}}]{lucek04b}
Lucek, E.~A., T.~S. Horbury, A.~Balogh, I.~Dandouras, and H.~R\`{e}me, Cluster
  observations of hot flow anomalies, \textit{J. Geophys. Res.}, \textit{109},
  A06,207, \doi{10.1029/2003JA010016}, 2004.

\bibitem[{\textit{Maynard et~al.}(2007)\textit{Maynard, Burke, Ober, Farrugia,
  Kucharek, Lester, Mozer, Russell, and Siebert}}]{maynard07}
Maynard, N.~C., W.~J. Burke, D.~M. Ober, C.~J. Farrugia, H.~Kucharek,
  M.~Lester, F.~S. Mozer, C.~T. Russell, and K.~D. Siebert, Interaction of the
  bow shock with a tangential discontinuity and solar wind density decrease:
  {O}bservations of predicted fast mode waves and magnetosheath merging,
  \textit{J. Geophys. Res.}, \textit{112}, A12,219, \doi{10.1029/2007JA012293},
  2007.

\bibitem[{\textit{McFadden et~al.}(2008a)\textit{McFadden, Carlson, Larson,
  Ludlam, Abiad, Elliott, Turin, Marckwordt, and Angelopoulos}}]{mcfadden08a}
McFadden, J.~P., C.~W. Carlson, D.~Larson, M.~Ludlam, R.~Abiad, B.~Elliott,
  P.~Turin, M.~Marckwordt, and V.~Angelopoulos, The {THEMIS} {ESA} plasma
  instrument and in-flight calibration, \textit{Space Sci. Rev.}, \textit{141},
  277--302, \doi{10.1007/s11214-008-9440-2}, 2008a.

\bibitem[{\textit{Omidi and Sibeck}(2007)}]{omidi07}
Omidi, N., and D.~G. Sibeck, Formation of hot flow anomalies and solitary
  shocks, \textit{J. Geophys. Res.}, \textit{112}, A01,203,
  \doi{10.1029/2006JA011663}, 2007.

\bibitem[{\textit{Omidi et~al.}(2010)\textit{Omidi, Eastwood, and
  Sibeck}}]{omidi10}
Omidi, N., J.~P. Eastwood, and D.~G. Sibeck, Foreshock bubbles and their global
  magnetospheric impacts, \textit{J. Geophys. Res.}, \textit{115}, A06,204,
  \doi{10.1029/2009JA014828}, 2010.

\bibitem[{\textit{Omidi et~al.}(2013a)\textit{Omidi, Zhang, Sibeck, and
  Turner}}]{omidi13shfa}
Omidi, N., H.~Zhang, D.~Sibeck, and D.~Turner, Spontaneous hot flow anomalies
  at quasi-parallel shocks: 2. {H}ybrid simulations, \textit{J. Geophys. Res.},
  \textit{118}, 173--180, \doi{10.1029/2012JA018099}, 2013a.

\bibitem[{\textit{Omidi et~al.}(2013b)\textit{Omidi, Sibeck, Blanco-Cano, D.,
  Turner, Zhang, and Kajdi\v{c}}}]{omidi13}
Omidi, N., D.~G. Sibeck, X.~Blanco-Cano, R.-C. D., D.~L. Turner, H.~Zhang, and
  Kajdi\v{c}, Dynamics of the foreshock compressional boundary and its
  connection to foreshock cavities, \textit{J. Geophys. Res.}, \textit{118},
  823--831, \doi{10.1002/jgra.50146}, 2013b.

\bibitem[{\textit{Paschmann et~al.}(1988)\textit{Paschmann, Haerendel, Sckopke,
  M{\"o}bius, L{\"u}hr, and Carlson}}]{paschmann88}
Paschmann, G., G.~Haerendel, N.~Sckopke, E.~M{\"o}bius, H.~L{\"u}hr, and C.~W.
  Carlson, Three-dimensional plasma structures with anomalous flow directions
  near the earth's bow shock, \textit{J. Geophys. Res.}, \textit{93}(A10),
  11,279--11,294, \doi{10.1029/JA093iA10p11279}, 1988.

\bibitem[{\textit{Plaschke et~al.}(2009)}]{plaschke09}
Plaschke, F., et~al., Statistical study of the magnetopause motion: First
  results from {THEMIS}, \textit{J. Geophys. Res.}, \textit{114}, A00C10,
  \doi{10.1029/2008JA013423}, 2009.

\bibitem[{\textit{Potemra et~al.}(1989)\textit{Potemra, Lühr, Zanetti,
  Takahashi, Erlandson, Marklund, Block, Blomberg, and Lepping}}]{potemra89}
Potemra, T.~A., H.~Lühr, L.~J. Zanetti, K.~Takahashi, R.~E. Erlandson, G.~T.
  Marklund, L.~P. Block, L.~G. Blomberg, and R.~P. Lepping, Multisatellite and
  ground-based observations of transient {ULF} waves, \textit{J. Geophys.
  Res.}, \textit{94}, 2543--2554, \doi{10.1029/JA094iA03p02543}, 1989.

\bibitem[{\textit{Schwartz}(1998)}]{schwartz98}
Schwartz, S.~J., \textit{Analysis Methods for Multi-Spacecraft Data}, chap.
  Shock and discontinuity normals, Mach numbers, and related parameters, pp.
  249--270, ISSI Scientific Reports SR-001, ESA Publications Division, 1998.

\bibitem[{\textit{Schwartz et~al.}(1988)\textit{Schwartz, Kessel, Brown,
  Woolliscroft, Dunlop, Farrugia, and Hall}}]{schwartz88}
Schwartz, S.~J., R.~L. Kessel, C.~C. Brown, L.~J.~C. Woolliscroft, M.~W.
  Dunlop, C.~J. Farrugia, and D.~S. Hall, Active current sheets near the
  earth's bow shock, \textit{J. Geophys. Res.}, \textit{93}(A10),
  11,295--11,310, \doi{10.1029/JA093iA10p11295}, 1988.

\bibitem[{\textit{Schwartz et~al.}(2000)\textit{Schwartz, Paschmann, Sckopke,
  Bauer, Dunlop, Fazakerley, and Thomsen}}]{schwartz00}
Schwartz, S.~J., G.~Paschmann, N.~Sckopke, T.~M. Bauer, M.~Dunlop, A.~N.
  Fazakerley, and M.~F. Thomsen, Conditions for the formation of hot flow
  anomalies at {E}arth's bow shock, \textit{J. Geophys. Res.}, \textit{105},
  12,639--12,650, \doi{10.1029/1999JA000320}, 2000.

\bibitem[{\textit{Schwartz et~al.}(2006)\textit{Schwartz, Sibeck, Wilber,
  Meziane, and Horbury}}]{schwartz06}
Schwartz, S.~J., D.~Sibeck, M.~Wilber, K.~Meziane, and T.~S. Horbury, Kinetic
  aspects of foreshock cavities, \textit{Geophys. Res. Lett.}, \textit{33},
  L12,103, \doi{10.1029/2005GL025612}, 2006.

\bibitem[{\textit{Schwartz et~al.}(1985)}]{schwartz85}
Schwartz, S.~J., et~al., An active current sheet in the solar wind,
  \textit{Nature}, \textit{318}, 269--271, \doi{10.1038/318269a0}, 1985.

\bibitem[{\textit{Shue et~al.}(2009)\textit{Shue, Chao, Song, McFadden,
  Suvorova, Angelopoulos, Glassmeier, and Plaschke}}]{shue09}
Shue, J.-H., J.-K. Chao, P.~Song, J.~P. McFadden, A.~Suvorova, V.~Angelopoulos,
  K.~H. Glassmeier, and F.~Plaschke, Anomalous magnetosheath flows and
  distorted subsolar magnetopause for radial interplanetary magnetic fields,
  \textit{Geophys. Res. Lett.}, \textit{36}, L18,112,
  \doi{10.1029/2009GL039842}, 2009.

\bibitem[{\textit{Shue et~al.}(1998)}]{shue98}
Shue, J.-H., et~al., Magnetopause location under extreme solar wind conditions,
  \textit{J. Geophys. Res.}, \textit{103}, 17,691--17,700,
  \doi{10.1029/98JA01103}, 1998.

\bibitem[{\textit{Sibeck et~al.}(2002)\textit{Sibeck, Phan, Lin, Lepping, and
  Szabo}}]{sibeck02}
Sibeck, D.~G., T.~D. Phan, R.~P. Lin, R.~P. Lepping, and A.~Szabo, Wind
  observations of foreshock cavities: A case study, \textit{J. Geophys. Res.},
  \textit{107}, SMP 4--1 -- SMP 4--10, \doi{10.1029/2001JA007539}, 2002.

\bibitem[{\textit{Sibeck et~al.}(1999)}]{sibeck99}
Sibeck, D.~G., et~al., Comprehensive study of the magnetospheric response to a
  hot flow anomaly, \textit{J. Geophys. Res.}, \textit{104}, 4577--4593,
  \doi{10.1029/1998JA900021}, 1999.

\bibitem[{\textit{Sonnerup and Scheible}(1998)}]{sonnerup98}
Sonnerup, B. U.~O., and M.~Scheible, \textit{Analysis Methods for
  Multi-Spacecraft Data}, chap. Minimum and maximum variance analysis, pp.
  185--220, ESA Publications Division, 1998.

\bibitem[{\textit{Thomas et~al.}(1991)\textit{Thomas, Winske, Thomsen, and
  Onsager}}]{thomas91}
Thomas, V.~A., D.~Winske, M.~F. Thomsen, and T.~G. Onsager, Hybrid simulation
  of the formation of a hot flow anomaly, \textit{J. Geophys. Res.},
  \textit{96}, 11,625--11,632, \doi{10.1029/91JA01092}, 1991.

\bibitem[{\textit{Thomsen et~al.}(1988)\textit{Thomsen, Gosling, Bame, Quest,
  Russell, and Fuselier}}]{thomsen88}
Thomsen, M.~F., J.~T. Gosling, S.~J. Bame, K.~B. Quest, C.~T. Russell, and
  S.~A. Fuselier, On the origin of hot diamagnetic cavities near the earth's
  bow shock, \textit{J. Geophys. Res.}, \textit{93}(A10), 11,311--11,325,
  \doi{10.1029/JA093iA10p11311}, 1988.

\bibitem[{\textit{Turner et~al.}(2011)\textit{Turner, Eriksson, Phan,
  Angelopoulos, Tu, Liu, Teh, McFadden, and Glassmeier}}]{turner11}
Turner, D.~L., S.~Eriksson, T.~D. Phan, V.~Angelopoulos, W.~Tu, W.~Liu, W.~L.
  Teh, J.~P. McFadden, and K.~H. Glassmeier, Multispacecraft observations of a
  foreshock-induced magnetopause disturbance exhibiting distinct plasma flows
  and an intense density compression, \textit{J. Geophys. Res.}, \textit{116},
  A04,230, \doi{10.1029/2010JA015668}, 2011.

\bibitem[{\textit{Turner et~al.}(2013)\textit{Turner, Omidi, Sibeck, and
  Angelopoulos}}]{turner13}
Turner, D.~L., N.~Omidi, D.~G. Sibeck, and V.~Angelopoulos, First observations
  of foreshock bubbles upstream of {E}arth's bow shock: Characteristics and
  comparisons to {HFA}s, \textit{J. Geophys. Res.}, \textit{118}, 1552--1570,
  \doi{10.1002/jgra.50198}, 2013.

\bibitem[{\textit{V\"{o}lk and Auer}(1974)}]{volk74}
V\"{o}lk, H.~J., and R.-D. Auer, Motions of the bow shock induced by
  interplanetary disturbances, \textit{J. Geophys. Res.}, \textit{79}, 40--48,
  \doi{10.1029/JA079i001p00040}, 1974.

\bibitem[{\textit{Wang et~al.}(2013a)\textit{Wang, Zong, and Zhang}}]{wang13a}
Wang, S., Q.~Zong, and H.~Zhang, Cluster observations of hot flow anomalies
  with large flow deflections: 1. velocity deflections, \textit{J. Geophys.
  Res.}, \textit{118}, 732--743, \doi{10.1002/jgra.50100}, 2013a.

\bibitem[{\textit{Wang et~al.}(2013b)\textit{Wang, Zong, and Zhang}}]{wang13b}
Wang, S., Q.~Zong, and H.~Zhang, Cluster observations of hot flow anomalies
  with large flow deflections: 2. bow shock geometry at {HFA} edges, \textit{J.
  Geophys. Res.}, \textit{118}, 418--433, \doi{10.1029/2012JA018204}, 2013b.

\bibitem[{\textit{Wu et~al.}(1993)\textit{Wu, Mandt, Lee, and Chao}}]{wu93}
Wu, B.~H., M.~E. Mandt, L.~C. Lee, and J.~K. Chao, Magnetospheric response to
  solar wind dynamic pressure variations: {I}nteraction of interplanetary
  tangential discontinuities with the bow shock, \textit{J. Geophys. Res.},
  \textit{98}, 21,297--21,311, \doi{10.1029/93JA01013}, 1993.

\bibitem[{\textit{Zhang et~al.}(2010)\textit{Zhang, Sibeck, Zong, Gary,
  McFadden, Larson, Glassmeier, and Angelopoulos}}]{zhang10}
Zhang, H., D.~G. Sibeck, Q.-G. Zong, S.~P. Gary, J.~P. McFadden, D.~Larson,
  K.-H. Glassmeier, and V.~Angelopoulos, Time {H}istory of {E}vents and
  {M}acroscale {I}nteractions during {S}ubstorms observations of a series of
  hot flow anomaly events, \textit{J. Geophys. Res.}, \textit{115}, A12,235,
  \doi{10.1029/2009JA015180}, 2010.

\bibitem[{\textit{Zhang et~al.}(2013)\textit{Zhang, Sibeck, Zong, Omidi,
  Turner, and Clausen}}]{zhang13}
Zhang, H., D.~G. Sibeck, Q.-G. Zong, N.~Omidi, D.~Turner, and L.~B.~N. Clausen,
  Spontaneous hot flow anomalies at quasi-parallel shocks: 1. {O}bservations,
  \textit{J. Geophys. Res.}, \textit{3357-3363}, 3357--3363,
  \doi{10.1002/jgra.50376}, 2013.

\end{thebibliography}

\end{article}

\clearpage{}

\clearpage{}

\clearpage{}

\end{document}